\documentclass[conference]{IEEEtran}
\usepackage{graphicx,amssymb,amsmath}%图片支持
\usepackage{multicol}%正文双栏
\usepackage{CJK}
\usepackage[noadjust]{cite}%参考文献
\usepackage{setspace}%调整行距
\usepackage{stfloats}
\usepackage{midfloat}
\usepackage{flushend,cuted}
\usepackage{bm}
\usepackage{multirow}
\usepackage{array}
\usepackage{stfloats}
\usepackage{cite}
\usepackage{algorithm}% http://ctan.org/pkg/algorithms
\usepackage{algpseudocode}% http://ctan.org/pkg/algorithmicx

\IEEEoverridecommandlockouts

\ifCLASSINFOpdf
\else
\fi
\newcommand{\E}{\textmd{E}}
\newcommand{\Var}{\textmd{Var}}
\newcommand{\diag}{\textmd{diag}}
\newcommand{\vect}{\textmd{vec}}
\newcounter{MYtempeqncnt}

\allowdisplaybreaks
\begin{document}
%\title{Energy-Efficient-Aware Relay Selection in Distributed Multipair Massive MIMO\\ Full Duplex Relaying System}
\title{Channel Estimation and Uplink Achievable Rates \\in One-Bit Massive MIMO Systems}

%\author{Yongzhi Li, Cheng Tao, Lingwen Zhang, Liu Liu%
%\IEEEoverridecommandlockouts
%\thanks{The research was supported in part by the National 863 Project Granted
%No. 2014AA01A706, the NSFC project under grant No. 61471027, the Fundamental
%Research Funds for the Central Universities under grant 2014JBZ001,
%the Research Fund of National Mobile Communications Research LaboratorySoutheast
%University (No. 2014D05), and Beijing Natural Science
%Foundation project under grant No.4152043.}%
%\thanks{Y. Li, C. Tao, L. Zhang and L. Liu are with the Institute of Broadband
%Wireless Mobile Communications, School of Electronic and Information
%Engineering, Beijing Jiaotong University, Beijing, 100044, China (email:
%liyongzhi@bjtu.edu.cn, chtao@bjtu.edu.cn,
%zhanglw@bjtu.edu.cn, liuliu@bjtu.edu.cn).}
%\thanks{C. Tao and L. Liu are also with the National Mobile Communications Research Laboratory,
%Southeast University.}
%}
\author{\IEEEauthorblockN{Yongzhi Li\IEEEauthorrefmark{1},
Cheng Tao\IEEEauthorrefmark{1},
Liu Liu\IEEEauthorrefmark{1},
Gonzalo Seco-Granados\IEEEauthorrefmark{2}, and
A. Lee Swindlehurst\IEEEauthorrefmark{3}}
\IEEEauthorblockA{\IEEEauthorrefmark{1} Institute of Broadband Wireless Mobile Communications, Beijing Jiaotong University, Beijing 100044, P.R.China.}
\IEEEauthorblockA{\IEEEauthorrefmark{2} Universitat Aut$\grave{\textrm{o}}$noma de Barcelona, Bellaterra, Barcelona 08193, Spain.}
\IEEEauthorblockA{\IEEEauthorrefmark{3} Center for Pervasive Communications and Computing (CPCC), University of California, Irvine, Irvine, CA 92697, USA.  \\
(Email: \{liyongzhi, chtao, liuliu\}@bjtu.edu.cn, gonzalo.seco@uab.es, swindle@uci.edu)}
%Telephone: (800) 555--1212, Fax: (888) 555--1212}
%\IEEEauthorblockA{\IEEEauthorrefmark{4}Tyrell Inc., 123 Replicant Street, Los Angeles, California 90210--4321}
}

\maketitle

\begin{abstract}
%\boldmath
This paper considers channel estimation and achievable rates for the uplink of a massive multiple-input multiple-output (MIMO) system where the base station is equipped with one-bit analog-to-digital converters (ADCs). By rewriting the nonlinear one-bit quantization using a linear expression, we first derive a simple and insightful expression for the linear minimum mean-square-error (LMMSE) channel estimator. Then employing this channel estimator, we derive a closed-form expression for the lower bound of the achievable rate for the maximum ratio combiner (MRC) receiver. Numerical results are presented to verify our analysis and show that our proposed LMMSE channel estimator outperforms the near maximum likelihood (nML) estimator proposed previously.
\end{abstract}
% if the conference you are submitting to favors bold math in the %abstract,
% of the abstract to achieve this. Many IEEE journals/conferences frown %on

%\begin{IEEEkeywords}
%Relay network, full duplex, massive MIMO, Rician fading.
%\end{IEEEkeywords}

% \fi
\IEEEpeerreviewmaketitle

\section{Introduction}
Massive multiple-input multiple-output (MIMO) communication systems are currently attracting significant research interest. Channel state information (CSI) plays an essential role in these systems, and it has been shown that, with CSI known at the base station (BS), simple signal processing techniques such as maximum-ratio combining (MRC) can be employed at the BS to reduce noise and interference among the terminals, and hence to significantly improve the spectral efficiency \cite{marzetta2010noncooperative, ngo2013energy}.

Most previous work has assumed that each antenna element and corresponding radio frequency (RF) chain is equipped with a high-resolution analog-to-digital converter (ADC). However, the power consumption of the ADCs grows exponentially with the number of quantization bits \cite{walden1999analog}, and power also grows with increased bandwidth and sampling rate requirements, as proposed in next generation systems. For massive MIMO configurations employing many antennas and ADCs, the cost and power consumption will be prohibitive, and alternative approaches are needed.

The use of low-cost one-bit ADCs is a potential solution to this problem.  One-bit ADCs consist of a simple comparator, they do not require automatic gain control or highly linear amplifiers, and hence they can be implemented with very low cost and power consumption \cite{ jianhua2015capacity, singh2009multi }. The authors of \cite{mezghani2008analysis,nossek2006capacity } showed that for one-bit ADCs, the capacity maximizing transmit signals for SISO channels are discrete, which is different from the infinite-resolution case where a Gaussian codebook is optimal. They also showed that the capacity of massive MIMO systems is not severely reduced by the coarse quantization at low signal-to-noise ratio (SNR). In fact, it has been shown in \cite{mezghani2007on} that the power penalty due to one-bit quantization is approximately equal to only $\pi/2$ (1.96dB) at low SNR. However, at high SNRs, one-bit quantization can produce a large capacity loss \cite{jianhua2014high}. In either case, the availability of accurate receiver-side CSI is indispensable for exploiting the full potential of massive MIMO systems, and an important open question is how to reliably estimate the channel and decode the data symbols when one-bit output quantization is employed. There has been related work on channel estimation and data detection with one-bit quantization in massive MIMO systems \cite{chiara2014massive,juncil2015near,jianhua2014channel,jacobsson2015one}. However, these methods rely on either the maximum-likelihood algorithm \cite{juncil2015near} or on iterative algorithms with high-complexity \cite{jianhua2014channel}. Moreover, the channel estimators and the achievable rate expressions obtained with these methods do not yield simple and insightful expressions.

%
%It is interesting that, although the output of the signal is quantized by a single bit, the rate loss in the low SNR region is not as large as we imagine. The authors in \cite{ nossek2006capacity } showed the rate loss due to the one-bit ADCs is 1.572bits/channel-use for $4\times 4$ MIMO with a 64QAM constellation. The authors in \cite{mezghani2007on} proved that the mutual information of MIMO with one-bit ADCs is only $2/\pi$ times smaller compared to that of the unquantized MIMO case in the low SNR region. Therefore, deploying the one-bit quantization ADCs at the BS can make the massive MIMO technique more feasible and viable in practice.

In this paper, we consider the uplink of a massive MIMO system with one-bit ADCs on each receive antenna, and we investigate the problem of channel estimation and determining the approximate achievable uplink rate. In particular, we provide a simple and insightful expression for the linear minimum mean-square-error (LMMSE) channel estimator for one-bit massive MIMO systems. Using this estimator and assuming an MRC receiver, we then obtain a simple closed-form lower bound for the achievable rate. Numerical results show that our proposed channel estimator outperforms the least squares (LS) and near maximum-likelihood (nML) channel estimators proposed in \cite{chiara2014massive,juncil2015near}.

%However, the one-bit quantization  generally has to tolerate large rate loss because of the severe threshold, especially in the high SNR region \cite{ jianhua2014high}. Therefore, since CSI is indispensable in massive MIMO system, an important question is how to reliably estimate the channel and decode the data symbols using one-bit output quantization.

\section{System Model}
We consider a single-cell one-bit massive MIMO system with $K$ single-antenna terminals and an $M$-antenna base station (BS). For uplink data transmission, the received signal at the BS is given by
\begin{equation}
{\mathbf{y}} = \sqrt {{\rho_d}}\, {\mathbf{Hs}} + {\mathbf{n}},
\end{equation}
% where $\mathbf{G}=\mathbf{H}\mathbf{D}^{1/2}\in\mathbb{C}^{M\times K}$ is the channel matrix between the BS and the $K$ users which models the independent fast fading, geometric attenuation, the log-normal shadow fading. We assume $\mathbf{H}\sim\mathcal{CN}(0,\mathbf{I})$ is the $M\times K$ matrix of fast fading coefficients between the $K$ users at the BS, and $\mathbf{D}=\diag\{\beta_1,...,\beta_K\}\in\mathbb{C}^{K\times K}$ is a diagonal matrix which denotes the large-scale fading of the $K$ users. To simply our representation, we ignore the large scale fading and assume that $\beta_k = 1$ for $\forall k$. In this case, we have $\mathbf{G} = \mathbf{H}$.
where $\mathbf{H}$ is the channel matrix between the BS and the $K$ users (with entries modeled as  zero-mean unit-variance complex Gaussian variables), $\sqrt{\rho_d} \mathbf{s}\in\mathbb{C}^{K\times 1}$ represents the data symbols simultaneously transmitted from the $K$ users, with $\E\{|s_k|^2\} = 1$ so that $\rho_d$ represents the average transmitted power of each user. The term $\mathbf{n}\sim\mathcal{CN}(0,\mathbf{I})\in\mathbb{C}^{M\times 1}$ denotes additive white Gaussian noise.%, whose entries are independent identically distributed zero-mean circularly symmetric complex random variables.

%By using one-bit quantization ADCs which are equipped at the BS,
The quantized signals
%\footnote{Throughout the whole paper, the {\it{quantized}} signal means the signal obtained after the ADCs.}
obtained after the one-bit ADCs are represented as
\begin{equation}\label{sysmtem_quantization}
\mathbf{r} = \mathcal{Q}(\mathbf{y}),
\end{equation}
where $\mathcal{Q}(.)$ represents the one-bit quantization operation, which is applied separately to the real and imaginary parts of the signal. The outcome of the one-bit quantization thus lies in the set $\mathcal{R} = 1/\sqrt 2 \{1+1j, 1-1j, -1+1j, -1-1j\}$, which includes without loss of generality a scaling factor so that the power of each quantized signal is one. %In this case, the power of the output will equal to $2$. In this paper, in order to normalize the power of the output, we assume the output set is $\mathcal{R} = \{1+1j, 1-1j, -1+1j, -1-1j\}/\sqrt{2}$.
Although $\mathcal{Q}(.)$ is obviously a  nonlinear operation, we can express $\mathbf{r}$ as
\begin{equation}\label{Bussgang}
\mathbf{r} = \mathbf{Ay}+\mathbf{q},
\end{equation}
where $\mathbf{A}\in\mathbb{C}^{M\times M}$ and $\mathbf{q}\in\mathbb{C}^{M\times 1}$ is the quantization noise. There are an infinite number ways of defining $\mathbf{A}$ and $\mathbf{q}$ for~(\ref{Bussgang}) to hold; in the next section we will present a common approach based on the Bussgang decomposition \cite{Bussgang:1952yq}.
\section{Channel Estimation in One-Bit Massive MIMO}
The authors in \cite{chiara2014massive,juncil2015near,jacobsson2015one,jianhua2014channel} have proposed various methods for channel estimation, relying on either the maximum-likelihood algorithm or iterative techniques. However, the channel estimators obtained by these methods do not yield much insight into the problem. In what follows, we will use the linear expression in \eqref{Bussgang} to derive a simple expression for the linear minimum mean-square-error (LMMSE) channel estimator.

\subsection{Channel Estimation}
In a practical system, the channel $\mathbf{H}$ has to be estimated at the BS, and it is used to detect the data symbols transmitted from the $K$ users. In the uplink transmission phase, we assume that the channel coherence interval is divided into two parts: one dedicated to training and the other to data transmission.

In the training stage, all $K$ users simultaneously transmit their pilot sequences of $\tau$ symbols each to the BS, which yields
\begin{equation}\label{received_signal_training}
\mathbf{Y}_p = \sqrt{\rho_p}\mathbf{H}\bm{\Phi}^T +\mathbf{N}_p,
\end{equation}
where $\mathbf{Y}_p\in\mathbb{C}^{M\times\tau}$ is the received signal, $\rho_p$ is the transmit power of each pilot symbol, and $\bm{\Phi}\in\mathbb{C}^{\tau\times K}$ is the matrix of pilot symbols.  To simplify the analysis, we assume orthogonal pilot sequences with $\tau=K$, {i.e.,} $\bm{\Phi}^T\bm{\Phi}^* = \tau \mathbf{I}$.  While this choice has been shown to be optimal for full-resolution ADCs \cite{hassibi2003how}, we recognize this may not be true for one-bit ADCs and leave this problem for future work.

Vectorizing the received signal yields
\begin{equation}\label{vec_received_signal_training}
\vect(\mathbf{Y}_p) = \mathbf{y}_p = \left(\bm{\Phi} \otimes \sqrt{\rho_p}\mathbf{I}_{M}\right)\underline{\mathbf{h}} + \mathbf{n}_p,
\end{equation}
where $\underline{\mathbf{h}} = \vect(\mathbf{H})$ and $\mathbf{n}_p = \vect(\mathbf{N}_p)$.
After the one-bit ADCs and using \eqref{Bussgang}, the quantized signal can be expressed as
\begin{equation}\label{Training_Quantized_Signal}
  \mathbf{r}_p = \mathcal{Q}(\mathbf{y}_p) = \tilde{\bm{\Phi}}\underline{\mathbf{h}} + \tilde{\mathbf{n}}_p,
\end{equation}
where the $i$th element of $\mathbf{r}_p$ takes values from the set $\mathcal{R}$, $\tilde{\bm{\Phi}} = \mathbf{A}_p\left(\bm{\Phi} \otimes \sqrt{\rho_p}\mathbf{I}_{M}\right) \in\mathbb{C}^{M\tau\times M\tau}$,  $\tilde{\mathbf{n}}_p =\mathbf{A}_p\mathbf{n}_p + \mathbf{q}_p\in\mathbb{C}^{M\tau\times 1}$, $\mathbf{A}_p\in\mathbb{C}^{M\tau\times M\tau}$ is a certain square matrix, and $\mathbf{q}_p\in\mathbb{C}^{M\tau\times 1}$ is the quantization noise.

According to \eqref{Training_Quantized_Signal}, we can readily see that the quantizer noise $\mathbf{q}_p$ is related to the matrix $\mathbf{A}_p$. A particularly meaningful choice for $\mathbf{A}_p$ is the one that minimizes the power of the quantizer noise $\mathbf{q}_p$ or, equivalently, that yields $\mathbf{q}_p$ uncorrelated with $\mathbf{y}_p$, as in \cite{Mezghani:2012rt}. This value of $\mathbf{A}_p$ is the result of
\begin{equation}\label{optimal_A_p_Orig}
\mathop {\arg} \min \limits_{\bf{A}_p} \E\{||\mathbf{q}_p||_2^2\} =\mathop {\arg} \min \limits_{\bf{A}_p} \E\{||\mathbf{r}_p - \mathbf{A}_p\mathbf{y}_p||_2^2\},
\end{equation}
whose solution is is given by
\begin{equation}\label{optimal_A_p}
\mathbf{A}_p = \mathbf{C}_{\mathbf{y}_p\mathbf{r}_p}^{H}\mathbf{C}_{\mathbf{y}_p\mathbf{y}_p}^{-1},
\end{equation}
where $\mathbf{C}_{\mathbf{y}_p\mathbf{r}_p}$ denotes the cross-correlation matrix between the received signal $\mathbf{y}_p$ and the quantized signal $\mathbf{r}_p$, and $\mathbf{C}_{\mathbf{y}_p\mathbf{y}_p}$ denotes the auto-correlation matrix of the received signal $\mathbf{y}_p$. For one-bit quantization and Gaussian signals, $\mathbf{C}_{\mathbf{y}_p\mathbf{r}_p}$ is given by \cite{Bussgang:1952yq}\cite[Ch. 10]{papoulis2002probability}
\begin{equation}\label{Cross_Quantize_Receive}
{{\bf{C}}_{{\mathbf{y}_p\mathbf{r}_p}}} = \sqrt {\frac{2}{\pi }} {{\bf{C}}_{{\mathbf{y}_p\mathbf{y}_p}}}{\rm{diag}}{\left( {{{\bf{C}}_{{\mathbf{y}_p\mathbf{y}_p}}}} \right)^{ - \frac{1}{2}}},
\end{equation}
where ${\rm{diag}}\left(\bf{X}\right)$ is a diagonal matrix formed from the diagonal elements of $\bf{X}$.
%Substituting \eqref{Cross_Quantize_Receive} into \eqref{optimal_A_p}, we have
%\begin{equation}
%  {{\bf{A}}_p} = \sqrt {\frac{2}{\pi }} {\rm{diag}}{\left( {{{\bf{C}}_{{\bf{YY}}}}} \right)^{ - \frac{1}{2}}}
%\end{equation}

Substituting \eqref{Cross_Quantize_Receive} into \eqref{optimal_A_p}, we have
\begin{align}\label{A_p1}
{{\bf{A}}_p} &= \sqrt {\frac{2}{\pi }} {\rm{diag}}{\left( {{{\bf{C}}_{{\mathbf{y}_p\mathbf{y}_p}}}} \right)^{ - \frac{1}{2}}} \nonumber\\
& = \sqrt {\frac{2}{\pi }} \diag\left(\left(\bm{\Phi}\bm{\Phi}^H \otimes \rho_p\mathbf{I}_{M}\right) + \mathbf{I}_{MK}\right)^{-\frac{1}{2}} \nonumber \\
& = \sqrt {\frac{2}{\pi } \frac{1}{1+K\rho_p}} \mathbf{I}_{MK}=\alpha_p\mathbf{I}_{MK}.
\end{align}
According to \cite[Ch. 12]{kay1993fundamentals}, the LMMSE channel estimate of $\underline{\mathbf{h}}$ is thus given by
\begin{equation}\label{MMSE_Channel}
\hat{\underline{\mathbf{h}}}^{\texttt{LM}} = \mathbf{C}_{\underline{\mathbf{h}}\mathbf{r}_p} \mathbf{C}_{\mathbf{r}_p\mathbf{r}_p}^{-1}\mathbf{r}_p,
\end{equation}
%\left(\tilde{\bm{\Phi}}\tilde{\bm{\Phi}}^H+\mathbf{C}_{\tilde{\mathbf{n}}_d\tilde{\mathbf{n}}_d}\right)^{-1}
where $\mathbf{C}_{\underline{\mathbf{h}}\mathbf{r}_p}$ is the cross-correlation matrix between $\underline{\mathbf{h}}$ and $\mathbf{r}_p$, and $\mathbf{C}_{\mathbf{r}_p\mathbf{r}_p}$ is the auto-correlation matrix of $\mathbf{r}_p$.

%Note that, in order to obtain the matrix $\mathbf{A}_p$, we only need to calculate the diagonal entries of $\mathbf{C}_{\mathbf{YY}}$, which means the average power of each receiving antenna. We denote $\bar{\mathbf{h}}_m$ and $\bar{\mathbf{n}}_{m,p}$ as the $m$th row of the channel matrix $\mathbf{H}$ and the AWGN matrix $\mathbf{N}_p$, respectively. Thus, the average power of $m$th antenna can be expressed as
%\begin{align}\label{power_antenna}
%&{\rm{E}}\left\{ {\left( {{\rho _p}{{{\bf{\bar h}}}_m}{{\bf{\Phi }}^T}{{\bf{\Phi }}^*}{\bf{\bar h}}_m^H + {{{\bf{\bar n}}}_m}{{{\bf{\bar n}}}_m^H}} \right)/K } \right\}\nonumber\\
%&= K\rho_p+1
%\end{align}
%
%Therefore, according to \eqref{A_p1} and \eqref{power_antenna}, we can obtain $\mathbf{A}_p = \alpha_p \mathbf{I}$, where $\alpha_p = \sqrt{2/(\pi K\rho_p+\pi)}$.
%
%After decorrelation and power normalization of the quantized signals, the BS can obtain the channel observations of all the users. Specifically, for the $k$th user, the BS obtains the channel observation as
%\begin{align}\label{r_k}
%{{\bf{r}}_{p,k}} &= {\bf{R}_p}\bm{\phi}_k^*/K {\rho _p}\nonumber\\
% &= \alpha_p{{\bf{h}}_k} + {{{\bf{\tilde q}}}_p}
%\end{align}
%where ${{\bf{\tilde q}}_p} = \left( {{{\bf{A}}_p}{{\bf{N}}_p}\bm{\phi}_k^* + {\bf{Q}}\bm{\phi}_k^*} \right)/\tau {\rho _p}$ can be regarded as the effective noise.

%It is interesting to note that the matrix $\mathbf{A}_p$ we derived in \eqref{optimal_A_p} is aligned with the Bussgang theorem. In this case, we can not only minimize the quantizer noise, but also can obtain the uncorrelated quantizer noise.

Note that since the elements of $\mathbf{y}_p$ are uncorrelated with $\mathbf{C}_{\mathbf{y}_p\mathbf{y}_p}=(K\rho_p+1)\mathbf{I}$, then according to the Bussgang theorem \cite{Bussgang:1952yq}, the elements of $\mathbf{r}_p$ are also uncorrelated and their auto-correlation matrix is $\mathbf{C}_{\mathbf{r}_p\mathbf{r}_p}=\mathbf{I}_{MK}$.  In fact, setting $\mathbf{A}_p = \alpha_p \mathbf{I}$ according to the Bussgang theorem, the quantization noise $\mathbf{q}_p$ is not only uncorrelated with the received signal $\mathbf{y}_p$ but also the channel $\underline{\mathbf{h}}$. Therefore, the LMMSE channel estimator can be obtained as
\begin{align}\label{MMSE_Channel_Approx}
\hat{\underline{\mathbf{h}}}^{\texttt{LM}} = \tilde{\bm{\Phi}}^H\mathbf{r}_p.
\end{align}

% We will show in the Section V that the relative gap between the \eqref{MMSE_Channel} and \eqref{MMSE_Channel_Approx} is rather small. Therefore, we use the \eqref{MMSE_Channel_Approx} to pursue a detailed analysis of the MSE behavior and the lower bound of the achievable rate in the rest of this paper .

%\begin{align}
%\mathbf{C}_{\mathbf{r}_p\mathbf{r}_p} %&= {\rm{E}}\left\{ \left( \mathbf{A}_p\mathbf{n}_p + \mathbf{q}_p \right)\left( \mathbf{A}_p\mathbf{n}_p + \mathbf{q}_p \right)^H \right\}\nonumber\\
%& = \E\left\{\left(\tilde{\bm{\Phi}}\mathbf{h} + \tilde{\mathbf{n}}_p\right)\left(\tilde{\bm{\Phi}}\mathbf{h} + \tilde{\mathbf{n}}_p\right)^H\right\} \nonumber\\
%& = \tilde{\bm{\Phi}}\tilde{\bm{\Phi}}^H + \mathbf{A}_p\mathbf{A}_p^H + \mathbf{C}_{\mathbf{q}_p \mathbf{q}_p}
%\end{align}
%with
%\begin{equation}
%{{\bf{C}}_{{\mathbf{q}_p\mathbf{q}_p}}} = {{\bf{C}}_{{\mathbf{r}_p\mathbf{r}_p}}} - \alpha_q^2{{\bf{C}}_{{\mathbf{y}_p\mathbf{y}_p}}}
%\end{equation}
%Therefore, the MMSE channel estimate of ${{\bf{h}}_k}$ can be rewritten as
%\begin{equation}
%{{\bf{\hat h}}_k} = \beta_q\mathbf{h}_k+\frac{1}{K\rho_p}\tilde{\mathbf{q}}_p
%\end{equation}
%Therefore, we have
%\begin{equation}
%{{\bf{C}}_{{\mathbf{q}_p\mathbf{q}_p}}} = \left( {1 - 2/\pi } \right){\bf{I}}_{MK}
%\end{equation}

\subsection{MSE of the Channel Estimate at High SNR}
%In this subsection, we are focusing on the MSE behavior of the LS and MMSE channel estimation in high SNR region. The MSE of the channel estimate can be expressed as
%In this subsection, we focus on the MSE behavior of the LMMSE channel estimator in high SNR region.
The normalized mean-squared error (MSE) of a given channel estimate $\hat{\underline{\mathbf{h}}}$ can be expressed as
\begin{equation}
{\cal M} = \rm{E}\left\{ \left\| \hat{\underline{\mathbf{h}}} - \underline{\mathbf{h}} \right\|_2^2 / \left\| \underline{\mathbf{h}}\right\|_2^2\right\}.
\end{equation}
For the LMMSE channel estimate \eqref{MMSE_Channel_Approx}, we have
\begin{align}\label{MSE_MMSE}
{{\cal M}}^{\texttt{LM}} &= {\rm{tr}}\left( \mathbf{I}_{MK} - \tilde{\bm{\Phi}}^H \tilde{\bm{\Phi}} \right)  /MK\nonumber\\
 &= 1 - \frac{2K\rho_p}{\pi K\rho_p +\pi}.
\end{align}
Thus, for high SNR $\rho_p\rightarrow\infty$, we have
\begin{equation}\label{Limit_MSE_MMSE}
\mathop {\lim }\limits_{{\rho _p} \to \infty } {{\cal M}} = 1 - \frac{2}{\pi}.
\end{equation}
From \eqref{Limit_MSE_MMSE} we see that, due to the effect of the quantization, the accuracy of the channel estimate cannot be reduced to zero by increasing the training power without limit.

%For the LS channel estimation case, the channel estimate can be expressed by using \eqref{r_k}, which is shown as
%\begin{equation}\label{LS}
%  {\bf{\hat h}}_k^{LS} = \alpha_p{{\bf{h}}_k} + {{\bf{\tilde q}}_p}
%\end{equation}
%Thus, the MSE of the LS channel estimate can be given by
%\begin{align}\label{MSE_LS}
%{{\cal M}^{LS}} &= {{{\rm{E}}\left\{ {\left\| {{\bf{\hat h}}_k^{LS} - {{\bf{h}}_k}} \right\|_F^2} \right\}}}/{{{\rm{E}}\left\{ {\left\| {{{\bf{h}}_k}} \right\|_F^2} \right\}}}\nonumber\\
% &= {\rm{tr}}\left( {(1-\alpha_p)^2\mathbf{I} + {{\bf{C}}_{{{{\bf{\tilde q}}}_p}{{{\bf{\tilde q}}}_p}}}} \right)/M
%\end{align}
%As we can see from \eqref{MSE_LS} that, in the high SNR region, i.e., $\rho_p\rightarrow\infty$, both $\alpha_p$ and $\tr\left(\mathbf{C}_{\tilde{\mathbf{q}}_p\tilde{\mathbf{q}}_p}\right)$ will converge to zero, thus we can have
%\begin{equation}\label{Limit_MSE_LS}
%\mathop {\lim }\limits_{{\rho _p} \to \infty } {{\cal M}^{LS}} = 1
%\end{equation}

%From \eqref{Limit_MSE_LS} and \eqref{Limit_MSE_MMSE} we can see that, MMSE is better than the LS in the high SNR region. In addition, for both LS and MMSE channel estimations, we cannot increase the accuracy of the channel estimate by only increasing the training power, since the MSEs of LS and MMSE channel estimation will converge to a fixed value in the high SNR region.

\section{Achievable Rate Analysis in One-Bit Massive MIMO Systems}
\subsection{MRC Receiver}
In the data transmission stage, we assume that the $K$ users simultaneously transmit their data symbols, represented as vector $\mathbf{s}$, to the BS. After the one-bit quantization, the signal at the BS can be expressed as
\begin{align}
\mathbf{r} _d &= \mathcal{Q}(\mathbf{y}_d)=\mathcal{Q}(\mathbf{H}\mathbf{s} +  \mathbf{n}_d) \nonumber \\
&=\sqrt{\rho_d}\mathbf{A}_d \mathbf{H}\mathbf{s} + \mathbf{A}_d\mathbf{n}_d + \mathbf{q}_d,
\end{align}
where the same definitions as in previous sections apply, but with the subscript $p$ replaced with $d$.  Following the same reasoning as in \eqref{optimal_A_p_Orig}-\eqref{A_p1}, in order to minimize the quantization noise (or equivalently, to make it uncorrelated with $\mathbf{y}_d$), $\mathbf{A}_d$ is chosen as $\mathbf{A}_d= \alpha_d \mathbf{I}$, with $\alpha_d = \sqrt{2/\pi (K \rho_d + 1)}$.

Next, we assume that the BS employs the MRC receiver to detect the data symbols transmitted by the $K$ users. Using a channel estimate $\hat{\underline{\mathbf{h}}}$, the quantized signal is separated into $K$ streams by multiplying it with a matrix $\hat{\mathbf{{H}}} = \vect^{-1}(\hat{\underline{\mathbf{h}}})$:
\begin{align}\label{MRC}
{\bf{\hat s}} &= {{{\hat{\mathbf{ {H}}}}}^H}\mathbf{r}_d\nonumber\\
&= \sqrt{\rho_d}{{{\hat{\mathbf{ {H}}}}}^H}{{\bf{A}}_d}(\hat{\mathbf{H}}\mathbf{s} + \bm{\mathcal{E}}\mathbf{s}) + {{{\hat{\mathbf{ {H}}}}}^H}{{\bf{A}}_d}{\bf{n}}_d + {{{\hat{\mathbf{ {H}}}}}^H}{\bf{q}}_d \; ,
\end{align}
where $\bm{\mathcal{E}} = \mathbf{H}-\hat{\mathbf{H}}$ denotes the channel estimation error. The $k$th element of $\hat{\mathbf{s}}$ is used to decode the signal transmitted from the $k$th user:
\begin{align}\label{hat_s_k}
  {\hat s_k} &= \underbrace {\sqrt {{\rho _d}} \hat{\mathbf{{h}}}_k^H{{\bf{A}}_d}{\hat{\bf{h}}_k}{s_k}}_{{\rm{Desired ~Signal}}}+ \underbrace {\sqrt {{\rho _d}} \hat{\mathbf{{h}}}_k^H\mathop \sum \limits_{i \ne k}^K {{\bf{A}}_d}{\hat{\bf{h}}_i}{s_i}}_{{\rm{User~Interference}}} \nonumber\\
  &+ \underbrace{\sqrt {{\rho _d}} {\mathbf{{w}}}_k^T \sum_{i =1}^K {{\bf{A}}_d}\bm{\varepsilon}_i{s_i}}_{{\rm{Estimate ~Eror}}} + \underbrace {\hat{\mathbf{{h}}}_k^H{{\bf{A}}_d}{\bf{n}}_d}_{{\rm{AWGN~Noise}}} + \underbrace {\hat{\mathbf{{h}}}_k^H{\bf{q}}_d}_{{\rm{Quant.~Noise}}},
\end{align}
where $\hat{\mathbf{{h}}}_k$ and $\bm{\varepsilon}_k$ are the $k$th columns of $\hat{\mathbf{{H}}}$ and $\bm{\mathcal{E}}$, respectively. Note that for a fixed number of users $K$, $\mathbf{A}_p$ and $\mathbf{A}_d$ are different only if $\rho_p \ne \rho_d$.

\subsection{Achievable Rate Analysis}
Although a number of previous papers have obtained expressions for the mutual information or the achievable rate of one-bit systems by using the joint probability distribution of the transmitted and received symbols \cite{jianhua2015capacity,chiara2014massive}, this approach does not result in easily computable or insightful expressions. To overcome this drawback, in this section we provide a closed-form expression for a lower bound of the achievable rate.

While quantization noise is not Gaussian distributed in general, a lower bound for the achievable rate can be found by modeling the quantization noise as Gaussian, since the Gaussian case corresponds to the worst case additive noise that minimizes the input-output mutual information \cite{hassibi2003how}.  In particular, the lower bound of the achievable rate can be obtained by modeling the quantization noise $\mathbf{q}_d$ as white Gaussian noise with the same covariance matrix:
\begin{equation}
\mathbf{C}_{\mathbf{q}_d\mathbf{q}_d} = \mathbf{C}_{\mathbf{r}_d\mathbf{r}_d} - \mathbf{A}_d\mathbf{C}_{\mathbf{y}_d\mathbf{y}_d}\mathbf{A}_d^H.
\end{equation}

%From the aspect of the information theorem and according to \eqref{hat_s_k},
Thus, the ergodic achievable rate of the uplink transmission in one-bit massive MIMO systems is lower bounded by \eqref{ergodic_achievable_rate}, shown on the next page \cite{hassibi2003how}.
\begin{figure*}[!t]
\normalsize
\setcounter{MYtempeqncnt}{\value{equation}}
\setcounter{equation}{19}
\begin{equation}\label{ergodic_achievable_rate}
\tilde{R}_k = \E\left\{\log_2\left(1+\frac{\rho_d \left|\hat{\mathbf{{h}}}_k^H{{\bf{A}}_d}{\hat{\bf{h}}_k}\right|^2}{\rho_d\sum_{i \ne k}^K\left|\hat{\mathbf{{h}}}_k^H{{\bf{A}}_d}{\hat{\bf{h}}_i}\right|^2 +  \rho_d\sum_{i =1}^K\left|\hat{\mathbf{{h}}}_k^H{{\bf{A}}_d}{\bm{\varepsilon}_i}\right|^2 + \left\|\hat{\mathbf{{h}}}_k^H{{\bf{A}}_d}\right\|^2 + \hat{\mathbf{{h}}}_k^H \mathbf{C}_{\mathbf{q}_d\mathbf{q}_d}\hat{\mathbf{{h}}}_k }\right)\right\}
\end{equation}
\hrulefill
\vspace*{-0.3cm}
\end{figure*}
In order to obtain a closed-form expression for the achievable rate, we first rewrite \eqref{hat_s_k} as a constant gain (which only depends on the channel distribution instead of the instantaneous channel) times the desired symbol plus an effective noise:
\begin{equation}
\setcounter{equation}{21}
{\hat s_k} = {\rm{E}}\left\{ {\sqrt {{\rho _d}} \hat{\mathbf{{h}}}_k^H{{\bf{A}}_d}{{\bf{h}}_k}} \right\}{s_k} + {{{\tilde n}}_{d,k}},
\end{equation}
where $\tilde{{n}}_{d,k}$ is the effective noise given by
\begin{align}
{{\tilde n}}_{d,k} =& \left ( {\sqrt {{\rho _d}} \hat{\mathbf{{h}}}_k^H{{\bf{A}}_d}{{\bf{h}}_k} - {\rm{E}}\left\{ {\sqrt {{\rho _d}} \hat{\mathbf{{h}}}_k^H{{\bf{A}}_d}{{\bf{h}}_k}} \right\}} \right){s_k} \nonumber\\
&+ \sqrt {{\rho _d}} \hat{\mathbf{{h}}}_k^H\mathop \sum \limits_{i \ne k}^K {{\bf{A}}_d}{{\bf{h}}_i}{s_i} + \hat{\mathbf{{h}}}_k^H{{\bf{A}}_d}{\bf{n}}_d + \hat{\mathbf{{h}}}_k^H{\bf{q}}_d . \label{eqnoise}
\end{align}% where $\mathbf{C}_{\mathbf{q}_d\mathbf{q}_d}$ is the covariance matrix of the quantizer noise $\mathbf{q}_d$, given by

%Note that, owing to the one-bit quantization ADC, the quantizer noise is not distributed as Gaussian. However, according to \cite{hassibi2003how}, the entropy of the effective noise is upper-bounded by the entropy of Gaussian noise. Therefore, we can obtain the lower bound of the achievable rate by using the fact that the worst-case additive noise is independent Gaussian noise of the same variance.

\textit{Lemma 1:} In a massive MIMO system with one-bit quantization, $\mathbf{A}_d=\alpha_d\mathbf{I}$, and the achievable rate is lower bounded by
\begin{equation}\label{Achievable_Rate}
R_k^{lb} = \log_2 \left( 1 + \frac{\rho_d\alpha_d^2\left| {\E\left\{ {\hat{\mathbf{{h}}}_k^H  {\mathbf{h}_k}} \right\}} \right|^2}{\rho_d\alpha_d^2\Var\left({\hat{\mathbf{{h}}}_k^H{{\bf{h}}_k}}\right) + \textrm{UI}_k + \textrm{AQN}_k }\right),
\end{equation}
where
\begin{equation}
\textrm{UI}_k = \rho_d\alpha_d^2\sum_{i\neq k }^K \E\left\{\left|\hat{\mathbf{{h}}}_k^H \mathbf{h}_i\right|^2\right\} ,
\end{equation}
\begin{equation}
\textrm{AQN}_k = (\alpha_d^2+1-2/\pi)\E\left\{\left\|\hat{\mathbf{{h}}}_k^H \right\|^2\right\} ,
\end{equation}
and $\Var(x)$ denotes the variance $x$.
%\begin{equation}
%\textrm{QN}_k = \E\left\{\left|\hat{\mathbf{{h}}}_k^H \mathbf{q}_d\right|^2\right\}
%\end{equation}

Using Lemma 1, we can provide a lower bound for the achievable rate with MRC processing in the following theorem.
%\begin{IEEEproof}
%See detail proof in Appendix A.
%\end{IEEEproof}

\textit{Theorem 1:} For MRC detection based on the LMMSE channel estimator, a lower bound for the achievable rate of a massive MIMO system with one-bit ADCs is given by
\begin{equation}\label{MRC_Achievable_Rate}
R_k^{lb} = {\log _2}\left( {1 + \frac{{{\rho _d}\alpha _d^2\alpha _p^2 K {\rho _p}M}}{{{\rho _d}\alpha _d^2 K + \alpha _d^2 + \left( {1 - 2/\pi } \right)}}} \right).
\end{equation}

\begin{IEEEproof}
The proof is omitted due to lack of space.
\end{IEEEproof}

\begin{figure}[!t]
  \centering
  \includegraphics[width=7.45cm]{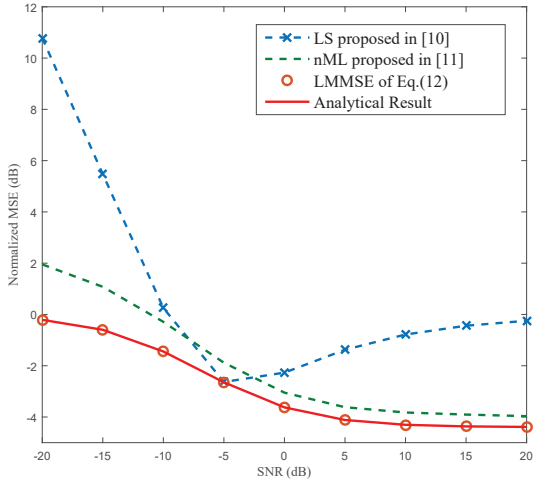}\\
  \vspace{-0.45cm}
  \caption{MSE of different channel estimators versus SNR with $M = 128$, $K = 8$. The Least Squares (LS) estimator is from \cite{chiara2014massive} and the near Maximum Likelihood (nML) estimator is from \cite{juncil2015near}.}\label{CE_LS_nML_MMSE}
  \vspace{-0.5cm}
\end{figure}

The lower bounds in \eqref{Achievable_Rate} and \eqref{MRC_Achievable_Rate} are obtained by approximating the effective noise $\tilde{{n}}_{d,k}$ as Gaussian. Since the effective noise is a sum of many terms (and especially for massive MIMO systems, where a large number of individual noise components form part of \eqref{eqnoise}), the central limit theorem provides confidence that this is a good approximation.
%However, an interesting question is: Is our derived achievable rate expression accurate enough to predict the system performance? To answer this question, we compare our achievable rate with the ergodic achievable rate, which is given as
%\begin{equation}\label{ergodic_achievable_rate}
%\footnotesize
%\begin{aligned}
%\hspace{-0.5cm}
%&\tilde{R}_k =  \\
% &\E\left\{\log_2\left(1+\frac{\rho_d \left|\hat{\mathbf{{h}}}_k^H{{\bf{A}}_d}{{\bf{h}}_k}\right|^2}{\E\left\{\left|\hat{\mathbf{{h}}}_k^H\left(\sqrt{\rho_d}\sum_{i \ne k}^K {{\bf{A}}_d}{{\bf{h}}_i}s_i + {{\bf{A}}_d}{\bf{n}}_d + {\mathbf{q}}_d\right)\right|^2\right\}}\right)\right\}
% \end{aligned}
%\end{equation}
In the next Section, we will show that the relative performance gap between \eqref{ergodic_achievable_rate} and the achievable rate lower bound given in Theorem 1 is small, which implies that the expression in \eqref{MRC_Achievable_Rate} is an excellent predictor of the system performance.

\section{Numerical Results}
For the simulations, we consider a massive MIMO system with one-bit ADCs, $M = 128$ BS antennas and $K=8$ users.
%Moreover, we assume the users transmit QPSK symbols to the BS both in training and data transmission stage. Therefore, the training matrix $\bm{\Phi}$ is composed of rows of Walsh-Hadamard matrix.
Fig.~\ref{CE_LS_nML_MMSE} compares the MSE of our proposed LMMSE channel estimator with the LS \cite{chiara2014massive} and the nML estimator \cite{juncil2015near}. The line "Analytical Result" is obtained using \eqref{MSE_MMSE}. We can see that our proposed LMMSE estimator outperforms the other two estimators along the entire SNR range. In addition, the MSE of the LMMSE estimator is aligned with our analytical results, which implies that our analysis is accurate.

\begin{figure}[!t]
  \centering
  \includegraphics[width=7.45cm]{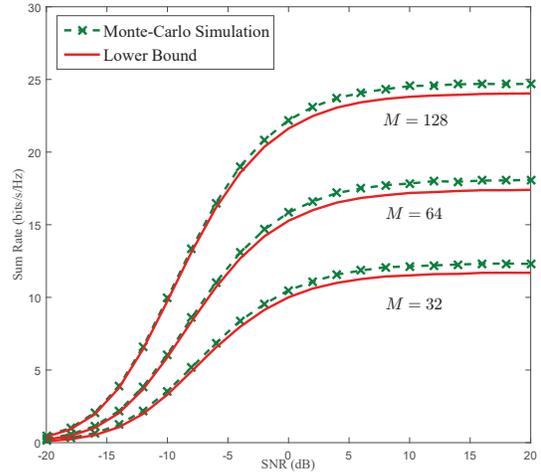}\\
  \vspace{-0.45cm}
  \caption{Sum Rate versus SNR with $K = 8$ and $\rho_p = \rho_d = \rm{SNR}$.}\label{SumRate_MRC}
  \vspace{-0.5cm}
\end{figure}

Next, we compare the closed-form lower bound of the achievable rate given in \eqref{MRC_Achievable_Rate} with the one given in \eqref{ergodic_achievable_rate}. Fig.~\ref{SumRate_MRC} shows the sum rate versus SNR with different numbers of receive antennas $M =\{32, 64 , 128\}$. This result confirms that the analytical expression given in Theorem 1 provides an excellent approximation to the known expression \eqref{ergodic_achievable_rate}, which has to be evaluated numerically.  Note that the agreement is especially tight for low SNRs, which is the expected operating region in massive MIMO.

% It is obvious that the relative performance gap between the cases with instantaneous CSI and statistical CSI is small, especially in the low SNR region. This implies that using the mean of the effective channel gain for signal detection is fairly reasonable, and the achievable rate given in \eqref{MRC_Achievable_Rate} is a good predictor of the system performance.

\section{Conclusions}
This paper has investigated channel estimation and achievable rates for massive MIMO systems with one-bit quantization. By rewriting the one-bit quantizer input-output relationship using a linear decomposition based on the Bussgang decomposition, we have derived a simple and insightful LMMSE channel estimator. Numerical results have shown that the proposed LMMSE channel estimator outperforms previously derived least-squares and the near maximum-likelihood channel estimators proposed in the literature. Then, using the proposed channel estimator, we have derived a closed-form expression for a lower bound on the achievable rate assuming the base station employs an MRC receiver. The gap between the lower bound and the ergodic achievable rate, which can be computed only numerically, is very small, and hence we can use our simplified result to accurately predict the system performance.
%\appendices

\section*{Acknowledgment}
The research was supported in part by the National 863 Project Granted No. 2014AA01A706, the Beijing Nova Programme (No. xx2016023), Beijing Natural Science Foundation project Grant No. 2016023 and No.4152043, the NSFC project under grant No. 61471027, the Research Fund of National Mobile Communications Research Laboratory, Southeast University No. 2014D05. A.~Swindlehurst was supported by the National Science Foundation under Grant ECCS-1547155, and by the Technische Universit\"at M\"unchen Institute for Advanced Study, funded by the German Excellence Initiative and the European Union Seventh Framework Programme under grant agreement No. 291763, and by the European Union under the Marie Curie COFUND Program.

\bibliographystyle{IEEEtran}
\bibliography{reference}

\end{document}